\documentclass[twocolumn,preprintnumbers,
  amsmath,amssymb]{revtex4}
  \usepackage{graphicx}
  \usepackage{xcolor}

\begin{document}
\title{ 
Langevin equations with non-Gaussian thermal noise: Valid but superfluous}
\author{Alex V. Plyukhin}
\email{aplyukhin@anselm.edu}
 \affiliation{ Department of Mathematics,
Saint Anselm College, Manchester, New Hampshire 03102, USA 
}

\date{\today}

\begin{abstract}
We discuss the statistics of additive thermal (internal) noise in systems governed by the generalized Langevin equation with linear dissipation.
To assess the equation's validity, it is common  to assume that the system is ergodic and to verify that solutions approach correct equilibrium values at asymptotically long times. In this paper, we instead consider the consistency 
of the generalized Langevin equation with the Jarzynski equality at finite times  and  do not assume the system's ergodicity. Specifically, we consider a classical Brownian oscillator whose initial stiffness,  or  frequency, 
is perturbed by a rectangular pulse of duration $\tau$. 
We find that the Jarzynski equality is  satisfied unconditionally only up to the seventh order in $\tau$; in higher orders, the Jarzynski equality holds \textit{if and only if} the noise is Gaussian. These results imply that, unless it is exact, the Langevin equation can only be used to evaluate properties that are linear or quadratic in noise and its derivatives. Such properties are insensitive to the noise statistics, so the Langevin equation with linear dissipation and non-Gaussian noise (though not inconsistent by itself) is superfluous.


\end{abstract}


\maketitle

\section{Introduction}

It is overwhelmingly common to assume that  noise $\xi(t)$ in the generalized Langevin equation (GLE)~\cite{Zwanzig,Kubo,Weiss} 
\begin{eqnarray}
   m\,\ddot q(t)=
   -\frac{\partial V(q,t)}{\partial q}-m\int_{0}^t \!\!
   \Gamma(t-t')\,\dot q(t')\,dt'+\xi(t)
   \label{gle0}
\end{eqnarray}
is a Gaussian stochastic process.
In this equation, $q$ and $m$ are the coordinate and the mass of a Brownian particle (a metaphor for an open mesoscopic system), $V(q,t)$ is the external potential, 
and 
the integral term represents the dissipation force, connected 
to zero-centered noise $\xi(t)$ by the fluctuation-dissipation relation (FDR)
\begin{eqnarray}
\langle \xi(t)\,\xi(t')\rangle= m T \,\Gamma(t-t').
\label{fdr}
\end{eqnarray}
The Gaussianity of $\xi(t)$ means that  higher order (multi-time) correlations of $\xi(t)$ are expressed in terms of pair (two-time) correlations. Since pair correlations are determined by the FDR, 
Eqs.~(\ref{gle0}) and (\ref{fdr}) characterize the random process $q(t)$ completely. This is fortunate and convenient, but looks like more of a fluke than a well-justified model. 

The Gaussianity of thermal noise is usually justified by observing that noise can be represented as the sum of $N$ random contributions, where $N$ is large; by the central limit theorem, the noise is Gaussian in the limit $N\to\infty$. However, this line of reasoning suggests that Gaussian noise is just an idealized  limiting case and invites us to consider deviations from Gaussian statistics.

Langevin equations with non-Gaussian thermal noise have been considered in a number of studies~\cite{Fox,Hanggi,Onuki,Grigolini,Caceres,Morgado,Sokolov,Speck,Klages,Jensen}. 
However, when solving the GLE with non-Gaussian noise, one immediately encounters a difficulty: 
at long times one recovers correct equilibrium values for the first two moments of $q$ and $v=\dot q$, but not for higher moments in general.
In contrast, for Gaussian noise  all the  moments relax to the correct equilibrium values, e.g.,
\begin{eqnarray}
    \langle v^2\rangle\to T/m, \,\,
    \langle v^4\rangle\to 3\,(T/m)^2,\,\, 
    \langle v^6\rangle\to 15\,(T/m)^3.
\label{equilibrium}
\end{eqnarray}
The proof of these relations is quite simple for delta-correlated noise and   in the absence of external potential~\cite{UO}, but it can be extended beyond those limitations. However, while the sufficiency of Gaussian noise to satisfy asymptotic conditions (\ref{equilibrium}) can be readily shown, its necessity, for the best of our knowledge, has never been proved for the GLE in the general form (\ref{gle0}).
Note also that the reasoning based on asymptotic relations (\ref{equilibrium}) does not apply to non-ergodic systems, which do not thermalize at long times~\cite{Plyukhin_NEO}.

Thus the situation appears to be somewhat puzzling.
On the one hand, the Gaussianity of noise is clearly sufficient for the consistency of the Langevin description at long times, at least for ergodic systems. On the other hand, 
it does not seem to follow from any general consideration based on the properties of the underlying dynamics. The Gaussianity of $\xi(t)$ is either introduced by hand (phenomenologically), or it emerges merely as a specific property of a special model. 


An example of the latter is the popular Caldeira-Leggett (CL) model~\cite{Zwanzig,Weiss} 
where the thermal bath is represented by independent oscillators in equilibrium linearly coupled to the system of interest. For that model one can derive GLE (\ref{gle0}) exactly with noise being a linear function of initial coordinates and velocities  of the equilibrium bath, which are Gaussian random variables. 
Therefore, for the CL model 
noise is strictly Gaussian. That, of course,  does not justify the Gaussianity of Langevin noise in general. On the contrary, 
the limitations (linearity) of the CL model suggest that in the general case noise is non-Gaussian. But that suggestion is apparently inconsistent with the observation that solutions of the GLE with non-Gaussian noise do not generally approach the correct equilibrium limit. 
Obviously, there is a contradiction between general physical expectations and mathematical properties  of the GLE.


Reimann~\cite{Reimann} suggested quite radical resolution of this contradiction. He argued that the statistics of noise $\xi(t)$ in the GLE (\ref{gle0}) does not depend on  specific details of the bath's composition, but  is uniquely determined by the form of the dissipation kernel $\Gamma(t)$ and temperature.  Then, since the CL model can produce (with a proper choice of coupling parameters) a GLE  with an arbitrary kernel $\Gamma(t)$, 
the statistics of Langevin noise for {\it any} model should be the same as for the CL  model, i. e. Gaussian. The GLE (\ref{gle0}) with non-Gaussian noise, according to Reimann, is intrinsically inconsistent. 

That conclusion appears to be in contradiction with a number of studies where the GLE with non-Gaussian noise was explicitly derived for models different than the CL model.  In particular, in Ref.~\cite{PS} we derived the GLE (\ref{gle0}) in the leading order in a relevant small parameter for a generalized Rayleigh model (where the bath is an ideal gas) and found that $\xi(t)$ is manifestly non-Gaussian. 
For a model with mode coupling the non-Gaussian statistics of Langevin noise was earlier demonstrated in Ref.~\cite{SO}.  A non-Gaussianity of Langevin noise is also implied, though indirectly, in   
Van Kampen's method of system size expansion~\cite{VK}. That method leads in higher perturbation orders to the master equation which differs from the Fokker-Planck equation, involving additional terms with derivatives of orders higher than two. That terms can be expressed in terms of higher-order cumulants of Langevin noise~\cite{OVK,Plyukhin3}, which signify the noise's non-Gaussianity.

In this paper we attempt to reconcile the conflicting points of view 
by analyzing affordances and  limitations of the GLE. The main message of the paper is as follows.  We argue that GLE (\ref{gle0}) with non-Gaussian noise is not inconsistent by itself,  as implied by Reimann. It is inconsistent, however, to use such an equation to evaluate properties involving moments and correlations of noise of orders higher than $2$.  But for such properties there is no differences between Gaussian and non-Gaussian noise. Therefore, GLE (\ref{gle0}) with non-Gaussian noise is not inconsistent, as suggested by Reimann, but merely redundant.


Now suppose one is interested in a quantity which requires noise correlations of orders higher than  $2$. As an example in this paper we consider the average exponential work $\langle e^{-W/T}\rangle$. For such quantity one cannot expect that the GLE (\ref{gle0}) with non-Gaussian noise  would generally give a correct result. One should either properly modify the equation, or limit oneself to regimes where noise correlations of higher orders are negligible and therefore the result does not depend on the noise statistics.

We illustrate that point by analyzing the consistency of the GLE with a fluctuation theorem, namely the Jarzynski equality, on a short times scale $\tau$. 
The example we consider might be interesting also in the context of the ongoing discussion about the consistency of non-Markovian Langevin models and fluctuation theorems (see Sec. II). Particularly relevant is the result  by Speck and Seifert~\cite{Speck}, which implies that the statistics of noise is not essential for non-Markovian Langevin models to recover fluctuation theorems.    In contrast, in this paper we found that  the GLE (\ref{gle0}) with non-Gaussian noise is consistent with a fluctuation theorem only to a few lowest orders in the process duration $\tau$, but not in general.

Note: In this paper we only address internal (thermal) noise, which is inherent in the combined system and its surrounding, or the thermal bath. The essential feature of internal noise, reflected in the FDR (\ref{fdr}), is that its intensity is proportional to the temperature of the bath. There is the large literature exploiting Langevin equations with both internal and external noises. 
External (or athermal) noise comes from sources outside of the system and the bath. It does not satisfy the FDR, and it 
is often considered to be non-Gaussian, see e. g.~\cite{Luczka,Baule,Kanazawa}. 
We do not consider external noise in this paper. As a tangential comment, there seems to be no  reason why external noise can be non-Gaussian and internal noise cannot.

\section{Limitations of GLE and Gaussianity}
It is important to distinguish the situations  when the GLE (\ref{gle0}) is exact and when it is approximate. Those cases, as we argue below, have to be handled differently.

For all known models where the GLE is exact, the noise is {\it exactly} Gaussian. That is the case of the CL and similar models, where equations of motion for the bath are linear and can be explicitly integrated. 
When exact, the GLE (\ref{gle0}) provides, of course, a complete and correct description.  In particular, it predicts the relaxation towards the correct equilibrium values at long times (for ergodic systems), see Eq.~(\ref{equilibrium}), and also, as discussed below, is consistent with fluctuation theorems on an arbitrary time scale.

The situation is quite different when the GLE (\ref{gle0}) is an approximate equation,  obtained perturbatively in the leading order of a relevant small parameter $\mu$. 
In the microscopic theory of Brownian motion~\cite{PS,Plyukhin3,MO,OVK,Mazo} the small parameter is the mass ratio $\mu^2=m_0/m$, where $m_0$ is the mass of bath molecules and $m\gg m_0$ is the mass of the Brownian particle.  Typically, one uses a projection operator technique to obtain an exact (``projected") equation of motion, which may (for the Zwanzig-Mori approach) look similar to the GLE (\ref{gle0}) but has obscure properties. In order to get a GLE  with desirable properties, in particular with $\xi(t)$ and $\Gamma(t)$ manifestly independent of the variables of the system, the exact projected equation has to be expanded in powers of $\mu$. As a result, one obtains in the lowest perturbation order the GLE (\ref{gle0}) with noise being of order $\mu$ and the dissipation kernel of order $\mu^2$, 
\begin{eqnarray}
    \xi(t)\sim \mu, \quad \Gamma(t)\sim \mu^2.
\end{eqnarray}
Note that this scaling is consistent with the FDR (\ref{fdr}).
Thus the equation turns out to be of order $\mu^2$, and therefore its
solutions are only meaningful to evaluate quantities of orders not higher than $\mu^2$. 
Since noise is linear in $\mu$, it only makes sense to consider pair (two-time) correlations of noise. But
under that restrictions, it  does not matter whether the noise Gaussian or not, since the differences appear only in correlations of 
higher orders.

Suppose, for instance, one is interested in the forth moment of the velocity $\langle v^4(t)\rangle$.
Solving the GLE, one can express 
$\langle v^4(t)\rangle$
as a functional of
the four-time noise correlation $\langle \xi(t_1)\,\xi(t_2)\,\xi(t_3)\,\xi(t_4)\rangle$. 
Since that correlation is of order $\mu^4$, 
we can be confident with the GLE's solution only if the equation is exact and noise is exactly Gaussian. 
If the GLE is not exact, but an approximate equation of order $\mu^2$ (with noise of an arbitrary statistics) we cannot expect it to give a meaningful result for  $\langle v^4(t)\rangle\sim\mu^4$. 
In order to manage the difficulty, the GLE (\ref{gle0}) should be abandoned and replaced 
with an appropriate generalized equation of order $\mu^4$. The latter  
involves an additional dissipation term nonlinear (cubic) in velocity~\cite{OVK,PS}. Another method is to derive a GLE immediately for the property of interest, in our example  directly for $v^4$, see Ref.~\cite{Albers,Plyukhin}. 

 
In order to illustrate and reinforce our point of view, in the rest of the paper we use the GLE to evaluate  the average exponential work $\langle e^{-W/T}\rangle$ for a certain process and assess conditions when the result satisfies the Jarzynski equality (JE)~\cite{J1,J2}, see also Appendix A.
We consider a specific cycle process of short duration $\tau$, for which the JE has (remarkably for any $\tau$) the form
\begin{eqnarray}
\left\langle e^{-W/T}\right\rangle=1.
\label{JE}
\end{eqnarray}
We use this equation as the criterion of the validity of the results obtained with the GLE. 
Ergodicity of the system is not required.
Evaluating the average exponential work perturbatively, 
we find that up to order $\tau^7$ it involves only pair correlations
of noise and its derivatives. As we argued above, for such correlations (of order $\mu^2$) the GLE (also of order $\mu^2$) is expected to give correct results regardless of the noise statistics. Indeed, we found explicitly that up to order $\tau^7$ the JE holds unconditionally. 
However, starting from order $\tau^8$ the average exponential work involves noise correlations of order higher than $2$. For such cases  we find that the JE holds if and only if noise is Gaussian.

There were many studies in recent years where the GLE with Gaussian noise was used to obtain, verify, and illustrate fluctuation theorems, and the JE in particular, see e.g. ~\cite{Zamponi,Mai,Ohta,Hasegawa,J3,Ritort}. 
Compared to those studies, our motivation in this paper is to some extent inverted: We adopt the JE as a
valid and rigorous result and consider constraints
which it imposes on the noise statistics.
 Speck and Seifert gave a proof of the JE for general non-Markovian processes emphasizing that  that noise may be non-Gaussian~\cite{Speck}. They used general arguments based on the existence and properties of a time-local (``substitute") operator which governs the evolution of the system's distribution function (though their demonstration of how to construct such an operator is limited to a Gaussian process only).
In contrast, our results show explicitly that  non-Markovian processes described by the GLE (\ref{gle0}) are consistent with the JE in general only if noise is Gaussian.

\section{Process}

Consider a classical Brownian oscillator in contact with
a single thermal bath
of temperature $T$, with the bare Hamiltonian
\begin{eqnarray}
    H(k)=\frac{p^2}{2m}+\frac{k(t)\, q^2}{2}.
\end{eqnarray}
The notation $H(k)$ is used to emphasize the parametric dependence of the Hamiltonian on the 
oscillator's stiffness $k(t)>0$, which is time-dependent and plays the role of  parameter $\lambda$ in the JE setting, see~\cite{J1,J2} and  Appendix A.
We shall refer to the oscillator as the system, and its surroundings as the bath.

The stiffness varies according to
the rectangular pulse protocol
\begin{eqnarray}
k(t) =
   \begin{cases}
     k_0, & t\le 0, \\
     k, & 0< t<\tau,\\
     k_0, & t\ge\tau.
   \end{cases}
   \label{k}
\end{eqnarray}
We denote the related frequencies as
\begin{eqnarray}
    \omega_0=\sqrt{k_0/m}, \quad \omega=\sqrt{k/m}.
\end{eqnarray}
For the brevity of notations we do not decorate 
$k$ and $\omega$, which are the values of $k(t)$ and $\omega(t)$ on the time interval $0< t<\tau$, with any subscript.  
We shall use the stiffness  values $k_0,k$ and the corresponding frequencies $\omega_0,\omega$ concurrently.


For $t<0$, the system and the bath are in thermal equilibrium with the distribution
\begin{eqnarray}
   \rho_0=Z_0^{-1}\,\exp(-H_0/T), \quad H_0=H(k_0). 
   \label{rho_0}
\end{eqnarray}
On the interval 
$0<t<\tau$, the combined system is out of equilibrium, and the oscillator is described  by the GLE (\ref{gle0}) with $V=m\,\omega^2q^2/2$,
\begin{eqnarray}
\ddot q(t)=-\omega^2\,q(t)-\int_{0}^t \Gamma(t-t')\,\dot q(t')\, dt'+\frac{1}{m}\,\xi(t).
\label{gle}
\end{eqnarray}
We assume that noise  $\xi(t)$
is stationary, zero-centered, $\langle \xi(t)\rangle=0$, and satisfies the fluctuation-dissipation relation, Eq.~(\ref{fdr}). 
We make no assumptions about a specific form of the dissipation kernel $\Gamma(t)$ and the statistics of $\xi(t)$.

The total work $W(\tau)$ on the system during the process is the sum of two contributions, $W(\tau)=W_1+W_2(\tau)$. The first contribution $W_1$ corresponds to the quench $k_0\to k$  and equals the change of the potential energy of the system
at time $t=0$,
\begin{eqnarray}
    W_1=\frac{k-k_0}{2}\, q^2,
\end{eqnarray}
where $q=q(0)$ is the system's coordinate at $t=0$. The second contribution $W_2$ corresponds to the reverse quench $k\to k_0$ and 
equals the change of the potential energy at time $t=\tau$,
\begin{eqnarray}
    W_2(\tau)=\frac{k_0-k}{2}\, q^2(\tau).
\end{eqnarray}
Then the total work equals
\begin{eqnarray}
    W(\tau)=\frac{k-k_0}{2}\, \left[q^2-q^2(\tau)\right].
    \label{work_gen}
\end{eqnarray}
Here the initial coordinate $q$ is distributed with distribution $\rho_0$, given by Eq.~(\ref{rho_0}), and $q(\tau)$ is to be found as a solution of the GLE (\ref{gle}).

Note that  Eq.~(\ref{work_gen}) can also be obtained from the general definition of the work~\cite{J1}
\begin{eqnarray}
    W=\int_{t_1}^{t_2}\dot \lambda(t)\, \frac{\partial H}{\partial \lambda}\, dt
\end{eqnarray}
replacing for the occasion  $t_1\to 0^-$, $t_2\to \tau^+$, and $\lambda(t)\to k(t)$, with
\begin{eqnarray}
    k(t)=k_0+(k-k_0)\,\theta(t)+(k_0-k)\,\theta(t-\tau),
  \end{eqnarray}  
and 
$\theta(t)$ be the unit-step function.

Since $k(0^-)=k(\tau^+)=k_0$, the process 
is cyclic, in which case the JE takes the form (\ref{JE}), see~\cite{J1,J2} and Appendix A.
Our goal is to assess conditions under which 
the solution of the GLE (\ref{gle}) is consistent with the JE.

The limiting case 
\begin{eqnarray}
  k(t)=k_0+a\,\delta (t),
  \label{delta}
\end{eqnarray}
which corresponds to $\tau\to 0$, $k-k_0\to\infty$, and $(k-k_0)\,\tau\to a$, was recently discussed in~\cite{Plyukhin2}. For that singular protocol we found that the JE is satisfied  regardless of the noise statistics.
One may argue, however, 
that the finding is trivial because in the limit $\tau\to 0$
the results  are governed merely by (Gaussian) statistics of the initial equilibrium state. Below we shall see that for finite 
$\tau$ the GLE (\ref{gle}) is consistent  with the JE (\ref{JE}) for any noise statistics only in lower orders in $\tau$. In order to satisfy the JE  in any order in $\tau$ noise in the GLE (\ref{gle}) must be Gaussian.

\section{Solving GLE}
In order to evaluate work with Eq.~(\ref{work_gen}) we need solution $q(t)$ of the GLE (\ref{gle}) on the time interval $0<t<\tau$.  The equation is linear and can be readily solved using Laplace transforms. The solution can be presented in the form
\begin{eqnarray}
    q(t)=q\,H(t)+v\, G(t)+\frac{1}{m}\,\{G*\xi\}(t).
    \label{sol}
\end{eqnarray}
Here $q=q(0)$ and $v=v(0)$ are the initial coordinate and velocity of the system, and the $G(t)$ and $H(t)$ are the relaxation functions with Laplace transforms (denoted by the tilde) 
\begin{eqnarray}
    \tilde G(s)&=&\frac{1}{s^2+s\,\tilde \Gamma(s)+\omega^2}, \nonumber\\ 
    \tilde H(s)&=&
  \frac{s+\tilde\Gamma}{s^2+s\,\tilde \Gamma+\omega^2}= \frac{1}{s}\,
  \left[1-\omega^2 \,\tilde G(s)\right].
\label{Green_def}
\end{eqnarray}
The last term in Eq.~(\ref{sol}) involves the convolution 
of $G(t)$ and noise $\xi$
\begin{eqnarray}
    \{G*\xi\}(t)=\int_0^t G(t-t')\, \xi(t')\, dt'.
    \label{convolution}
\end{eqnarray}
As follows from Eq. (\ref{Green_def}), the relaxation functions are related in the time space as
\begin{eqnarray}
 H(t)=1-\omega^2\! \int_0^t G(t')\, dt',  
 \label{H_time}
\end{eqnarray}
and their  initial values are
\begin{eqnarray}
    G(0)=0,\quad H(0)=1.
    \label{Green_initial}
\end{eqnarray}
Note that $H(t)$ is dimensionless and $G(t)$ has the dimension of time.


We shall also need the  derivatives of the relaxation functions. One can directly verify the validity of the following relation in the Laplace domain:
\begin{eqnarray}
    s\,\tilde G(s)= \tilde H(s)-\tilde G(s)\,\tilde\Gamma(s).
\end{eqnarray}
In the time domain it determines the derivative
of $G$, while the derivative of $H$ can be found from Eq.~(\ref{H_time}),
\begin{eqnarray}
    \dot G(t)&=&H(t)-\{G*\Gamma\}(t),\nonumber\\
    \dot H(t)&=&-\omega^2\,G(t).
    \label{Green_derivatives}
\end{eqnarray}

Note that noise appears in solution (\ref{sol})  in the form of the convolutions with $G$. 
To facilitate further calculations let us
introduce the dimensionless convolution 
\begin{eqnarray}
    \eta(t)=\alpha\,\{G*\xi\}(t),
    \qquad \alpha=\frac{1}{m}\sqrt{\frac{k_0}{T}}
    \label{eta}
\end{eqnarray}
and rewrite solution (\ref{sol}) as
\begin{eqnarray}
    q(t)=q\,H(t)+v\,G(t)+\sqrt{\frac{T}{k_0}}\,\eta(t).
    \label{sol1A}
\end{eqnarray}
The first two terms in this expression are deterministic, while $\eta(t)$ can be viewed as a zero-centered dimensionless noise. In what follows we shall refer to $\eta(t)$ as convoluted noise. Its statistical  properties are determined by those of Langevin noise $\xi(t)$, but also depend on the bare properties of the system.

\section{Work}
Combining Eq.~(\ref{work_gen}) for work $W$
and solution (\ref{sol1A}) for $q(t)$, one can present the dimensionless work
\begin{eqnarray}
    w(\tau)=W(\tau)/T
\end{eqnarray}
in the form
\begin{eqnarray}
    &&\hspace{-1.2cm}
    w=\frac{k-k_0}{2T}\,\Big\{q^2\left[1-H^2(\tau)\right]-v^2\,G^2(\tau)-\frac{T}{k_0}\,\eta^2(\tau)\nonumber\\
    &&\hspace{-1.2cm}
    -2 \sqrt{\frac{T}{k_0}}\,\eta(\tau)\, \big[q\, H(\tau)+v\,G(\tau)\big]-2\, q\, v\,G(\tau)\,H(\tau)
    \Big\}.
    \label{work1}
\end{eqnarray}
The randomness enters this expression through convoluted noise $\eta(t)$ and also through the initial variables  $q$ and $v$ of the system.  The latter are distributed according the canonical distribution $\rho_0$, 
\begin{eqnarray}
    \rho_0(q,v)=\frac{1}{Z_0}\,
    \exp\left\{ -\frac{m v^2}{2T}-\frac{k_0 q^2}{2T}  \right\}.
\end{eqnarray}
It is convenient to work with dimensionless coordinate $x$ and velocity $y$,
\begin{eqnarray}
    x=\frac{q}{q_0}, \quad y=\frac{v}{v_0},
\end{eqnarray}
where $x_0$ and $v_0$ are the root mean squares of the coordinate and velocity in equilibrium
\begin{eqnarray}
    q_0=\sqrt{\frac{T}{k_0}}, \quad v_0=\sqrt{\frac{T}{m}}.
\end{eqnarray}
Random variables $x$ and $y$ are distributed with the two-dimensional standard normal distribution
\begin{eqnarray}
    f_0(x,y)=\frac{1}{2\pi}\,\exp\left(-\frac{x^2+y^2}{2}\right),
    \label{f0}
\end{eqnarray}
have zero first moments and correlation, and unit  variances,
\begin{eqnarray}
  \langle x\rangle=\langle y\rangle=\langle x y\rangle=0, \quad
  \langle x^2\rangle=\langle y^2\rangle=1.
  \label{xy_moments}
\end{eqnarray}
The average of a dynamical variable $A$ over the system's equilibrium initial variables, which we denote as $\langle A\rangle_s$
\begin{eqnarray}
   \langle A\rangle_s=\int dq\,dv\,\rho_0(q,v)\, A(q,v),
\end{eqnarray}
can be written in terms of $x$ and $y$ as
\begin{eqnarray}
   \langle A\rangle_s=\int dx\,dy\,f_0(x,y)\, A(x,y). 
   \label{av_s}
\end{eqnarray}

By changing variables $(q,v)\to (x,y)$ in Eq.~(\ref{work1}), one can present  the dimensionless work 
as a quadratic form of  $x$ and $y$, 
\begin{eqnarray}
    w=a\,x^2+b\,y^2+c\,x\,y+d\,x+f\,y+g,
    \label{work2}
\end{eqnarray}
where the coefficients are dimensionless functions of the duration of the process $\tau$,
\begin{eqnarray}
    &&a=\frac{r}{2}\,\big[1-H^2(\tau)\big],\quad
    b=-\frac{r}{2}\,\big[\omega_0\, G(\tau)]^2, \nonumber\\
    &&c=-r\,H(\tau)\,\big[\omega_0\,G(\tau)\big],\quad
    d=-r\,H(\tau)\, \eta(\tau), \nonumber\\
    &&f=-r\,\big[\omega_0\, G(\tau)\big]\,\eta(\tau)\quad g=-\frac{r}{2}\,\eta^2(\tau),
\label{abc}
\end{eqnarray}
and parameter $r$ is defined as
\begin{eqnarray}
    r=\frac{k-k_0}{k_0}>-1.
    \label{r}
\end{eqnarray}
Note that coefficients $a,b,c$ are deterministic, while coefficients $d,f,g$ depend on convoluted noise $\eta(\tau)$. 

Although our goal is to evaluate the average exponential work $\langle e^{-w}\rangle$, we postpone that to the next section and first consider the average dimensionless work  
$\langle w\rangle$, which is often the quantity of interest.

First, we take the average of Eq.~(\ref{work2}) over initial variables of the system, i. e. 
over $x$ and $y$. Using Eq.~(\ref{xy_moments}), we find
$\langle w\rangle_s=a+b+g$, or more explicitly
\begin{eqnarray}
\langle w\rangle_s        
    =\frac{r}{2}\left[
    1-H^2(\tau)-\omega_0^2\, G^2(\tau)-\eta^2(\tau)
    \right].
\end{eqnarray}
In order to get the total average $\langle w\rangle$ we need to average the last term on the right-hand side over noise realizations,
\begin{eqnarray}
    \langle w\rangle        
    =\frac{r}{2}\left[
    1-H^2(\tau)-\omega_0^2\, G^2(\tau)-
    \langle \eta^2(\tau)\rangle
    \right].
    \label{work_av}
\end{eqnarray}
The evaluation of the variance of convoluted noise $\eta$
\begin{eqnarray}
 \langle \eta^2(\tau)\rangle=\frac{k_0}{m^2T}\,\langle (G*\xi)^2\rangle   
\end{eqnarray}
can be done using 
the fluctuation-dissipation relation (\ref{fdr}) and standard tricks of two-variable calculus,
\begin{eqnarray}
    \langle \eta^2(t)\rangle=
    \left(\frac{\omega_0}{\omega}\right)^2\,\left[1-H^2(t)\right]-\omega_0^2\, G^2(t).
\label{eta_variance}
\end{eqnarray}
Note that this expression is temperature-independent. Also, one can get from the above relation a useful constraint
\begin{eqnarray}
    1-H^2(t)-\omega^2\,G^2(t)> 0,
    \label{constraint}
\end{eqnarray}
which holds for any $t>0$.

Substituting Eq.~(\ref{eta_variance}) into Eq.~(\ref{work_av}) yields
\begin{eqnarray}
    \langle w(\tau)\rangle=\frac{1}{2}\,\frac{r^2}{r+1}\,\left[
 1-H^2(\tau)\right],
\label{work_av_result}
\end{eqnarray}
or
\begin{eqnarray}
    \langle w(\tau)\rangle=
    \frac{(k-k_0)^2}{2\, k_0\, k}\,\left[
    1-H^2(\tau)\right].
\label{work_av_result2}
\end{eqnarray}
Taking into account Eq.~(\ref{constraint}), one observes that 
result (\ref{work_av_result}) is consistent with the second law of thermodynamics which requires $\langle w(\tau)\rangle> 0$ for any $\tau>0$.

For the considered process,  the work is quadratic in noise, see Eq.~(\ref{work1}). The average work is determined by the two-time noise correlation and therefore 
does not depend on the statistics of noise. 
On the other hand,  any nonlinear function of work, and in particular the average exponential work 
$\langle e^{-w}\rangle$ discussed in the next sections, is expected to depend on the noise statistics.

\section{Exponential work: partial average}
The significance of the average exponential work 
$\langle e^{-w}\rangle$ is that it satisfies the JE, see Eqs.~(\ref{JE}) and (\ref{JE_A}).  We evaluate the average of the microscopic exponential work $e^{-w}$ in two steps. 
The first step, considered in this section, is to 
take the average over dimensionless initial coordinate $x$ and velocity $y$ of the system
\begin{eqnarray}
    \langle e^{-w}\rangle_s=\int dx\,dy\,f_0(x,y)\, e^{-w}, 
    \label{aux5}
\end{eqnarray}
see Eq.~(\ref{av_s}). Substituting here $w$ in the form (\ref{work2}) and denoting $z_1=x$ and $z_2=y$, one can present Eq.~(\ref{aux5}) as
\begin{eqnarray}
    \langle e^{-w}\rangle_s=e^{-g}\, I,
\end{eqnarray}
where $I$ is the two-dimensional Gaussian integral
\begin{eqnarray}
    I\!=\!\frac{1}{2\pi}\int dz_1\,dz_2\, \exp\Big[-\sum_{i,j=1}^2A_{ij} z_i z_j+\sum_{i=1}^2B_i\,z_i\Big],
    \label{I}
\end{eqnarray}
$A_{ij}$ and $B_i$ are the entries of the matrices 
\begin{eqnarray}
A=
\begin{pmatrix}
    a+1/2 & c/2\\
    c/2 & b+1/2
\end{pmatrix},
\quad
B=-\begin{pmatrix}
    d\\
    f
\end{pmatrix}
,
\end{eqnarray}
and $a,b,c,d,f,g$ are determined by Eq.~(\ref{abc}). The general result for Gaussian integral (\ref{I}) is~\cite{Balak}
\begin{eqnarray}
I=\frac{1}{2\sqrt{\det(A)}}\, \exp\left(
\frac{1}{4}\,\sum_{i,j=1}^2 A_{ij}^{-1} B_i B_j\right),
\end{eqnarray}
where $A_{ij}^{-1}$ are the entries of the
inverse matrix $A^{-1}$.
For our specific  $A$ we get $\det(A)=\Delta/4$
with
\begin{eqnarray}
    \Delta=4ab+2a+2b-c^2+1
    \label{Delta_def}
\end{eqnarray}
and 
\begin{eqnarray}
A^{-1}=\frac{4}{\Delta}
\begin{pmatrix}
    b+1/2 & -c/2\\
    -c/2 & a+1/2
\end{pmatrix}
.
\end{eqnarray}
Substituting Eq.~(\ref{abc}) for $a,b,c$ into Eq.~(\ref{Delta_def}) yields for $\Delta(t)$ a more explicit expression in terms of the relaxation functions
\begin{eqnarray}
    \Delta(t)=1+r\left[1-H^2(t)-\omega^2\,G^2(t)\right].
\label{Delta2}
\end{eqnarray}
Since parameter $r=(k-k_0)/k_0$ can take values in the interval $(-1,\infty)$, and the combination $1-H^2-\omega^2 G^2$ is positive, see Eq.~(\ref{constraint}), it follows from Eq.~(\ref{Delta2}) that $\Delta(t)$ is positive for any $t$. Interestingly, $\Delta(t)$ can also be expressed in terms of the variance of $\eta(t)$. From Eqs.~(\ref{r}),  (\ref{eta_variance}), and (\ref{Delta2}) one finds
\begin{eqnarray}
    \Delta(t)=1+r(r+1)\,\langle \eta^2(t)\rangle. 
\label{Delta3}
\end{eqnarray}

Combining the above results, taking into accounts Eq.~(\ref{abc}), and after some algebra, we obtain
\begin{eqnarray}
    \langle e^{-w(\tau)}\rangle_s=\frac{1}{\sqrt{\Delta(\tau)}}\,\exp\left[
    \frac{r(1+r)}{2}\,\frac{\eta^2(\tau)}{\Delta(\tau)}
    \right].
\label{expwork_s}
\end{eqnarray}
This result presents the exponential work averaged over (the initial variables of) the system. It is still a stochastic quantity governed by convoluted noise $\eta(t)$. Another function $\Delta(t)$ involved in Eq.~(\ref{expwork_s}) is deterministic. 

\section{Exponential work: total average}

In order to get the total average 
$\langle e^{-w}\rangle$ we need to 
average  expression (\ref{expwork_s}) over noise realizations. Recall that convoluted noise 
$\eta(t)$, defined by Eq.~(\ref{eta}), 
\begin{eqnarray}
    \eta(t)=\alpha\,\{G*\xi\}=\alpha\int_0^t G(t-t')\,\xi(t')\,dt',
\end{eqnarray}
is a functional of Langevin noise $\xi(t)$. Therefore, the average in question is over an ensemble of realizations of $\xi(t)$ on the time interval $(0,\tau)$. In general, such averaging requires a technique based on path integrals. Here,  as an alternative, we 
expand Eq.~(\ref{expwork_s}) in the Maclaurin series in powers of $\tau$
\begin{eqnarray}
     \langle e^{-w(\tau)}\rangle_s=1+c_1\,\tau+c_2\,\tau^2+\cdots. \label{aux1}
     \label{expansion0}
\end{eqnarray}
In this expansion coefficients 
\begin{eqnarray}
   c_n=\frac{1}{n!} \frac{d^n}{d\tau^n}\, \langle e^{-w(\tau)}\rangle_s
   \Big{|}_{\tau=0}
\label{c_general}
\end{eqnarray}
are (with the exception of $c_1,c_2,c_3$, see below)  stochastic functions which depend on initial values of the  derivatives of convoluted noise. The latter are expressed in terms of 
Langevin  noise $\xi(t)$, e. g.
\begin{eqnarray}
    \dot\eta(0)=0,\quad 
    \ddot\eta(0)=\alpha\,\xi(0), \quad
    \dddot\eta(0)=\alpha\,\dot\xi(0).
\end{eqnarray}
Although expressions for initial values of higher derivatives 
of $\eta(t)$
are more complicated (see Appendix B), it is clear that coefficients $c_n$ are determined by the statistics of Langevin noise $\xi(t)$ at the single moment $t=0$. 

Taking the average of expansion (\ref{expansion0}) over noise we get
\begin{eqnarray}
     \langle e^{-w(\tau)}\rangle=1+\langle c_1\rangle\,\tau+\langle c_2\rangle\,\tau^2+\cdots.
     \label{expansion}
\end{eqnarray}
By construction, this represents the total average of the exponential work. According to the JE (\ref{JE}), it equals $1$ for any $\tau$.  Therefore, in expansion (\ref{expansion}) all terms but the first  must vanish, i.e.
\begin{eqnarray}
 \langle c_n\rangle=0   
 \label{condition}
\end{eqnarray}
for $n\ge 1$.  This condition is expected to 
put constraints on the statistics of $\xi(t)$, and our goal is to find them.

We  evaluate several (non-averaged) coefficients $c_n$ using Eq.~(\ref{c_general}) with $\langle e^{-w}\rangle_s$ given by Eq.~(\ref{expwork_s}) and $\Delta(t)$ taken in the form of  
Eq.~(\ref{Delta2}).
The evaluation requires initial values of the derivatives of the relaxation functions $G(t)$ and $H(t)$ and convoluted noise $\eta(t)$. These values are calculated in Appendix B. The resulting 
explicit expressions for $c_n$ are very bulky, but can be readily handled with a symbolic computation software. 

We find that 
the first three coefficients do not depend on noise and vanish identically, 
\begin{eqnarray}
    c_1=c_2=c_3=0.
\end{eqnarray}

For $c_4$ we find
\begin{eqnarray}
    c_4=\frac{r}{8}\,\left[(1+r)\, \alpha^2\,\xi^2-\omega^2 \,\Gamma\right].
    \label{c4}
\end{eqnarray}
Here and below we use the notations
\begin{eqnarray}
   \xi=\xi(0), \,\,\dot\xi=\dot\xi(0),\,\, \Gamma=\Gamma(0), \,\,
   \dot\Gamma=\dot\Gamma(0),  \cdots
\end{eqnarray}
for initial values of noise, the dissipation kernel, and their derivatives.
From the definitions of $\alpha$, Eq.~(\ref{eta}), and $r$, Eq.~(\ref{r}), one finds $(1+r)\alpha^2=k/(m^2 T)$, while the FDR gives  $\langle \xi^2\rangle=m T \,\Gamma$. Taking that relations into account, we find from Eq.~(\ref{c4}) that the average $c_4$ vanishes,
\begin{eqnarray}
    \langle c_4\rangle=0.
\end{eqnarray}

For $c_5$ we obtain
\begin{eqnarray}
    c_5=\frac{r}{12}\,(1+r)\,\alpha^2\,\xi\,\dot \xi.
\end{eqnarray}
Differentiating the fluctuation-dissipation relation (\ref{fdr}) by one argument and then putting both arguments to $0$, one gets
\begin{eqnarray}
    \langle\xi\,\dot\xi\rangle=m\, T\,\dot\Gamma=0,
    \label{aux7}
\end{eqnarray} 
because for stationary noise $\Gamma(t)$ is an even function and therefore $\dot \Gamma=\dot\Gamma(0)=0$. Then the average $c_5$
vanishes too
\begin{eqnarray}
    \langle c_5\rangle=0.
\end{eqnarray}

 The expression for $c_6$   
 involves  $\xi^2,\,
 \dot\xi^2$, and $\xi\,\ddot\xi$. From the FDR one finds for the corresponding moments
\begin{eqnarray}
\langle \xi^2\rangle=m\, T\,\Gamma,\,\,
\langle \dot\xi^2\rangle=-m\, T\,\ddot\Gamma,\,\,
\langle \xi\,\ddot\xi\rangle=m\, T\, \ddot\Gamma.
    \label{aux3}
\end{eqnarray}
With that relations, we find that the average $c_6$ is zero,
\begin{eqnarray}
    \langle c_6\rangle=0.
\end{eqnarray}

The expression for $c_7$ involves 
$\xi\,\dot\xi$, $\dot\xi\ddot\xi$, and $\xi\dddot\xi$. The corresponding correlations, similarly to Eq.~(\ref{aux7}), are zero
\begin{eqnarray}
  \langle \xi\,\dot\xi\rangle=
  \langle \dot\xi\,\ddot\xi\rangle=
  \langle \xi\dddot\xi\rangle=0.
  \label{aux4}
\end{eqnarray}
As a result, the average $c_7$ vanishes
\begin{eqnarray}
    \langle c_7\rangle=0.
\end{eqnarray}

Summarizing the above calculations, 
one observes that the first three coefficients $c_1,c_2,c_3$ do not depend on noise, and the next four 
$c_4,c_5,c_6,c_7$
depend on noise only through pair products of noise $\xi=\xi^{(0)}$ and its derivatives
$\xi^{(n)}$, $1\le n\le 3$. Thus, the averaging of the first seven coefficients requires only 
pair correlations $\langle \xi^{(k)}\,\xi^{(n)}\rangle$, which are 
determined by the FDR,
and do not require the statistics of noise.
We therefore conclude that for the considered process the GLE is correct (consistent with the JE) up to order $\tau^7$ for an arbitrary noise statistics.

In orders higher than $\tau^7$ the situation is different and more general in the sense that average coefficients $\langle c_n\rangle$ with $n\ge 8$
involve not only pair correlations $\langle\xi^{(k)}\,\xi^{(n)}\rangle$, but also moments and correlations of higher orders, which cannot be determined from the FDR and depend on the noise statistics. 

The expression for $\langle c_8\rangle$, which is too bulky to present here, involves the forth moment 
$\langle\xi^4\rangle$. It also involves pair moments (\ref{aux4}) and in addition pair moments
\begin{eqnarray}
    \langle {\ddot\xi\,}^2\rangle=
    \langle \xi\,\xi^{(4)}\rangle=
    m\, T\,\Gamma^{(4)},\quad
    \langle\dot\xi\,\dddot\xi\rangle=-m\,T\,\Gamma^{(4)},
    \label{aux8}
\end{eqnarray}
which can be evaluated by differentiating the FDR.
Substituting Eqs.~(\ref{aux4}) and (\ref{aux8}) into the expression for $\langle c_8\rangle$, the latter is simplified to the form
\begin{eqnarray}
    \langle c_8\rangle=\frac{r^2 \omega^4}{128\,m^2 T^2}\,
    \Big\{
    \langle \xi^4\rangle-3\,m^2\,T^2\,\Gamma^2
    \Big\}.
\end{eqnarray}
We observe that $\langle c_8\rangle$ vanishes if and only if 
\begin{eqnarray}
    \langle \xi^4\rangle=3\,\langle\xi^2\rangle^2=3\,m^2\,T^2\,\Gamma^2,
\label{result1}
\end{eqnarray}
which is a property of the Gaussian distribution.

While suggestive, this finding, of course, is not yet a proof that $\xi(t)$ is a Gaussian process.
Condition (\ref{result1}) characterizes the random variable $\xi=\xi(0)$, i. e. the random process $\xi(t)$ at specific time $t=0$. Since noise is assumed to be stationary, the same condition holds for any times $t_i$. 
However, the Gaussianity of individual random variables $\xi(t_i)$ does not necessarily mean
that they are jointly Gaussian, that is, that an arbitrary set $\{\xi(t_1),\xi(t_2)\dots\xi(t_k)\}$ is described by a joint Gaussian multivariate distribution.
Therefore, even if the Gaussianity conditions like (\ref{result1}) hold for any moments $\langle\xi^n\rangle$, they do not guarantee that $\xi(t)$ is a Gaussian  process.
However, the analysis of coefficients $c_n$ of higher orders reveals that condition (\ref{condition}), $\langle c_n\rangle=0$, indeed requires the Gaussianity of the entire process $\xi(t)$, rather than just of a single random variable $\xi=\xi(0)$.  

For instance, for $\langle c_9\rangle$ we obtain, taking into account second moments like (\ref{aux7}), (\ref{aux3}) and (\ref{aux4}),
\begin{eqnarray}
    \langle c_9\rangle=\frac{r^2}{96}\,(1+r)^2\,\alpha^4\, \langle \xi^3\,\dot \xi\rangle.
\end{eqnarray}
In order to prove that  the moment $\langle \xi^3\,\dot \xi\rangle$ in this expression is zero, the FDR is not sufficient and we need to assume the Gaussianity of $\xi(t)$, which implies
\begin{eqnarray}
    &&\langle \xi(t_1)\,\xi(t_2)\,\xi(t_3)\,\xi(t_4)\rangle=\nonumber\\
    &&\quad\langle \xi(t_1)\,\xi(t_2)\rangle\,
    \langle \xi(t_3)\,\xi(t_4)\rangle\nonumber\\
    &&\quad \quad+
    \langle \xi(t_1)\,\xi(t_3)\rangle\,
    \langle \xi(t_2)\,\xi(t_4)\rangle\nonumber\\
    &&\quad\quad\quad +
     \langle \xi(t_1)\,\xi(t_4)\rangle\,
    \langle \xi(t_2)\,\xi(t_3)\rangle.
    \label{gauss4}
\end{eqnarray}
For equal times this property is reduced to Eq.~(\ref{result1}), but of course it is more restrictive than that. Using the FDR, we can re-write the above relation in terms of the dissipation kernel $\Gamma(t)$,
\begin{eqnarray}
    &&\langle \xi(t_1)\,\xi(t_2)\,\xi(t_3)\,\xi(t_4)\rangle=(m\,T)^2\Big\{\Gamma(t_1-t_2)\,\Gamma(t_3-t_4)\nonumber\\
    &&+\Gamma(t_1-t_3)\,\Gamma(t_2-t_4)+
\Gamma(t_1-t_4)\,\Gamma(t_2-t_3)\Big\}.
\label{gauss2}
\end{eqnarray}
Differentiating by, say, $t_1$, and then putting $(t_1,t_2,t_3,t_4)\to 0$, one finds
\begin{eqnarray}
    \langle \xi^3\,\dot \xi\rangle=3\, (m\,T)^2\,\Gamma\,\dot\Gamma=0,
    \label{aux11}
\end{eqnarray}
so $\langle c_9\rangle$ vanishes.

Next, the expression for $\langle c_{10}\rangle$ involves two forth moments, namely $\langle \xi^4\rangle$ and $\langle\xi^3\ddot\xi\rangle$.
Substituting for the former expression (\ref{result1}) we find
\begin{eqnarray}
    \langle c_{10}\rangle=\frac{r^2\,\omega^4}{384\,m^2 T^2}\,\Big\{
    \langle \xi^3\,\ddot\xi\rangle-3\, m^2 T^2 \Gamma\,\ddot\Gamma
    \Big\}.
\end{eqnarray}
Differentiating Eq.~(\ref{gauss2}) twice by $t_1$ and then putting all time arguments to $0$, we find
\begin{eqnarray}
    \langle \xi^3\,\ddot \xi\rangle=3\, (m\,T)^2\,\Gamma\,\ddot\Gamma,
    \label{aux12}
\end{eqnarray}
and therefore $\langle c_{10}\rangle=0$.

No new features are expected for coefficients of higher orders, so we conclude that the Gaussianity of $\xi(t)$ is sufficient to satisfy
condition (\ref{condition}) of the consistency of the GLE and the JE. 
Moreover, the following arguments show that 
the noise Gaussianity is not only sufficient for Eq.~(\ref{condition}) to hold
but also necessary.  Indeed, suppose that stochastic process $\xi(t)$ is non-Gaussian and yet satisfies Eq. (\ref{condition}). Then the right-hand side of Eq.~(\ref{gauss4}) for the forth moment  would contain an additional term $\kappa_4(t_1,t_2,t_3,t_4)$, which is the fourth cumulant of the process. In order to satisfy Eq.~(\ref{result1}), (\ref{aux11}), (\ref{aux12}) and also similar relations, which emerge for coefficients $c_n$ of higher orders, $\kappa_4$ and its partial derivatives of all orders at $t_1=t_2=t_3=t_4=0$
must be zero. That means,  assuming that function $\kappa_4$ is analytic,  that $\kappa_4$ is zero identically. Same arguments can be used to prove that cumulants of higher orders vanish as well and therefore $\xi(t)$ is Gaussian.

\section{Conclusion}
We have shown, albeit only for a specific process,  that the GLE is consistent with the JE if and only if the statistics of noise is Gaussian.
This does not mean the GLE with non-Gaussian noise is incorrect. But it does show that such equations can only be used to evaluate properties that are insensitive to noise statistics.

In the CL-like models, noise is a linear function of the equilibrium bath's initial variables, and the Gaussianity of the noise is merely a relic of the Gaussianity of that variables.  In the general case, i. e., when noise is not linear in bath's initial variables,
microscopic derivations lead to the  GLE (\ref{gle0}) with non-Gaussian noise. 
Derived perturbatively, such equations  are valid only to the second  order in a relevant small parameter $\mu$ and contain noise which is linear in $\mu$. Therefore, it is perturbatively inconsistent to use a GLE with non-Gaussian noise to evaluate properties involving the moments of noise of orders  higher than $2$. 

For the process of duration $\tau$ that we considered in this paper, the averaged exponential work 
is expressed up to order $\tau^7$
in terms of second moments  of noise and its derivatives, so the GLE is expected to give a correct description.
We confirmed this by showing that up to order $\tau^7$ the solution of the GLE is consistent with the JE regardless of the noise statistics. 

Beginning with order $\tau^8$, the average exponential work involves moments of noise of orders higher than $2$. It is described correctly by the GLE only if the equation is exact and noise is Gaussian (as for the CL model). 
Otherwise, the GLE fails to recover the JE and therefore does not describe the process correctly.

Thus, we argue that the GLE (\ref{gle0}) with non-Gaussian noise is not inconsistent per se, but can be exploited only when evaluating properties that are linear or quadratic in the noise. But that, of course, would be redundant because such properties do not depend on the noise statistics at all.  
That conclusion does not rely on assumptions about the system's thermalization at long times, and therefore applies to both ergodic and non-ergodic systems.
If one wishes to study the effects of the non-Gaussianity of noise, it is perturbatively  inconsistent to use a GLE of the form (\ref{gle0}). For that purpose, a more appropriate GLE would be of a higher perturbation order and would involve nonlinear dissipation terms~\cite{PS}.



\renewcommand{\theequation}{A\arabic{equation}}
  \setcounter{equation}{0}  

\section*{Appendix A: Jarzynski equality} 

Consider an open system whose bare Hamiltonian $H(\lambda)$
depends on a parameter $\lambda(t)$ which can be 
varied by an external agent.  In this paper 
the role of $\lambda(t)$ is played by the oscillator stiffness $k(t)$.
Suppose at $t<0$ parameter $\lambda$   has 
a fixed value $\lambda_0$ and the system is prepared in equilibrium with a thermal bath 
at temperature $T$ (in energy units), described by the canonical distribution 
\begin{eqnarray}
   \rho_0=Z_0^{-1}\,\exp(-H_0/T), \quad H_0=H(\lambda_0).
   \label{rho_0_A}
\end{eqnarray}
Next, during the finite time interval $0<t<\tau$, the parameter is changed from $\lambda_0=\lambda(0)$ to new value $\lambda_1=\lambda(\tau)$ according to a specified protocol
$\lambda(t)$. 
The quantity of interest is the work on the system $W$ required to arrange this process.
If the protocol is very slow,  the system 
at time $t=\tau$ is close 
a new equilibrium state with distribution  
\begin{eqnarray}
   \rho_1=Z_1^{-1}\,\exp(-H_1/T), \quad H_1=H(\lambda_1),
   \label{rho_1}
\end{eqnarray}
and the work takes the  (non-fluctuating) value $W=\Delta F$, where 
\begin{eqnarray}
    \Delta F=F_1-F_0=-T\,\ln \frac{Z_1}{Z_0}
\end{eqnarray}
is the difference of the Helmholtz free 
energy  $F$ in equilibrium
states corresponding to $\lambda_0$ and $\lambda_1$.
If $\tau$ is short, work $W$ strongly fluctuates, the system at $t=\tau$ is far from equilibrium, and 
distribution (\ref{rho_1}) appears to be irrelevant.
Yet the Jarzynski equality (JE) 
states that the average exponential work is still determined by 
the free energy difference~\cite{J1}
\begin{eqnarray}
\left\langle e^{-W/T}\right\rangle=e^{-\Delta F/T}.
\label{JE_A}
\end{eqnarray}
Surprisingly, the JE involves only initial and final values of $\lambda$ and does not depend on how fast and in what way that change is arranged.
Curiously, 
the JE involves the free energy $F_1$ of an equilibrium  state which is not involved in the process.  Note that the average in Eq.~(\ref{JE_A}) is taken over the equilibrium values of the combined system and the bath.

The proof of the JE was originally presented for systems weakly coupled to the thermal bath~\cite{J1}, but later it was extended for the strong coupling as well~\cite{J2}.


In this paper  we consider a cyclic process with $\lambda_0=\lambda_1$. In that case $\Delta F=0$ and the JE is reduced to Eq.~(\ref{JE}),
which is also known as the Bochkov-Kuzovlev relation~\cite{BK}

\renewcommand{\theequation}{B\arabic{equation}}
  \setcounter{equation}{0}  
\section*{Appendix B: Higher derivatives of $G,H,\eta$}
In order to evaluate coefficients $c_n$ in expansion (\ref{expansion0}) one needs initial values of derivatives of the relaxation functions $G(t),H(t)$ and of convoluted noise $\eta(t)$. In this appendix we evaluate initial values of derivatives up to order $10$, used in the main text.  
For all functions below, we use for initial values the notation $f=f(0)$. Derivatives of lower orders are denoted by dots and of  higher orders as $f^{(n)}$.

From Eqs.(\ref{Green_initial}), (\ref{Green_derivatives}) and (\ref{eta}) we find
\begin{eqnarray}
    &&G=0, \quad H=1, \quad \eta=0, \nonumber\\
    &&\dot G=1,\quad \dot H=0,\quad \dot\eta=0,\nonumber\\
   &&\hspace{-0.5cm}
   \ddot G=0,\quad \ddot H=-\omega^2,\quad \ddot\eta=\alpha\,\xi, \nonumber\\
   &&\hspace{-1cm}
   \dddot G=-\omega^2-\Gamma,\quad
   \dddot H=0,\quad
   \dddot\eta=\alpha\, \dot \xi,
\label{D03}
\end{eqnarray}
where $\alpha$ is defined by Eq.~(\ref{eta}).
Expressions for derivatives of higher orders involves initial values of the dissipation kernel $\Gamma$ and its derivatives $\Gamma^{(n)}$.  Since the noise $\xi(t)$ is assumed to be stationary, its autocorrelation function $\Gamma(t)$ is an even function. Therefore the initial values of derivatives of $\Gamma(t)$ of odd orders are zero,
\begin{eqnarray}
    \dot \Gamma=\dddot\Gamma=\Gamma^{(5)}=\cdots=0.
\end{eqnarray}
Taking that into account we find, by differentiating Eq.~(\ref{Green_derivatives}), the following (recurrence) relations
\begin{eqnarray}
&&G^{(4)}=0, \quad
H^{(4)}=-\omega^2\,\dddot G,\nonumber\\
&&\eta^{(4)}=\alpha\left(
\ddot\xi+\dddot G\,\xi\right)
\label{D4}
\end{eqnarray}
for derivatives of order $4$, 
\begin{eqnarray}
&&\hspace{-0.5cm}
G^{(5)}=H^{(4)}-\ddot\Gamma-\dddot G\,\Gamma, \quad
H^{(5)}=0,\nonumber\\
&&\hspace{0.5cm}
\eta^{(5)}=\alpha\left(
\dddot\xi-\dddot G\,
\dot\xi\right)
\label{D5}
\end{eqnarray}
for derivatives of order $5$, 
\begin{eqnarray}
    G^{(6)}=0, \quad 
    H^{(6)}=-\omega^2\,G^{(5)},\nonumber\\
    \eta^{(6)}=\alpha\left(
    \xi^{(4)}+\dddot G\,\ddot\xi+
    G^{(5)}\,\xi
    \right)
   \label{D6} 
\end{eqnarray}
for derivatives of order $6$, 
\begin{eqnarray}
    &&G^{(7)}=H^{(6)}\!-\!\Gamma^{(4)}\!-\!
    \dddot G\,\ddot\Gamma\!-\!G^{(5)}\, \Gamma, \nonumber\\
    &&H^{(7)}=0,\nonumber\\
    && \eta^{(7)}=\alpha\left(
    \xi^{(5)}+\dddot G\,\dddot\xi+
    G^{(5)}\,\dot\xi
    \right)
\label{D7}
\end{eqnarray}
for derivatives of order $7$, and 
\begin{eqnarray}
    &&G^{(8)}=0, \quad 
    H^{(8)}=-\omega^2\,G^{(7)},\nonumber\\
    && \eta^{(8)}=\alpha\left(
    \xi^{(6)}+\dddot G\,\xi^{(4)}+
    G^{(5)}\,\ddot\xi+G^{(7)}\,\xi
    \right)
\label{D8}
\end{eqnarray}
for the derivative of order $8$. We also use in the paper the expressions 
\begin{eqnarray}
&&G^{(9)}=H^{(8)}-\Gamma^{(6)}-\dddot G\,\Gamma^{(4)}-G^{(5)}\,\ddot\Gamma-G^{(7)}\,\Gamma,\nonumber\\
    &&H^{(9)}=0,
\end{eqnarray}
and $H^{(10)}=-\omega^2\,G^{(9)}$.


\end{document}